\documentclass[journal,article,submit,pdftex,moreauthors]{Definitions/mdpi} 
\firstpage{1} 
\pubvolume{1}
\issuenum{1}
\articlenumber{0}
\pubyear{2026}
\copyrightyear{2025}
\datereceived{ } 
\daterevised{ } 
\dateaccepted{ } 
\datepublished{ } 
\hreflink{https://doi.org/} 

\let\linenumbers\relax

\Title{\textit{Ab initio} study of the halo structure in $^{11}$Be}

\Author{Shihang Shen $^{1,2,*}$\orcidA{}, Serdar Elhatisari $^{3,4}$\orcidE{}, Dean Lee $^{6}$\orcidB{}, Ulf-G. Meißner $^{5,7,1,*}$\orcidD{} and Zhengxue Ren $^{8}$\orcidC{}}

\AuthorNames{Shihang Shen, Serdar Elhatisari, Dean Lee, Ulf-G. Meißner and Zhengxue Ren}

\address{%
$^{1}$ \quad Peng Huanwu Collaborative Center for Research and Education, International Institute for Interdisciplinary and Frontiers,
Beihang University, Beijing 100191, China \\
$^{2}$ \quad School of Physics, Beihang University, Beijing 102206, China \\
$^{3}$ \quad King Fahd University of Petroleum and Minerals (KFUPM), 31261 Dhahran, Saudi Arabia \\
$^{4}$ \quad Faculty of Natural Sciences and Engineering, Gaziantep Islam Science and Technology University, Gaziantep 27010, Turkey \\
$^{5}$ \quad Helmholtz-Institut für Strahlen- und Kernphysik and Bethe Center for Theoretical Physics, Universität Bonn, D-53115 Bonn, Germany \\
$^{6}$ \quad Facility for Rare Isotope Beams and Department of Physics and Astronomy, Michigan State University, East Lansing, Michigan 48824, USA \\
$^{7}$ \quad Institute for Advanced Simulation (IAS-4), Forschungszentrum Jülich, D-52425 Jülich, Germany \\
$^{8}$ \quad School of Physics, Nankai University, Tianjin 300071, China \\
}

\corres{Correspondence: sshen@buaa.edu.cn; meissner@hiskp.uni-bonn.de}

\abstract{
We present an \textit{ab initio} study of the one-neutron halo nucleus $^{11}$Be using nuclear lattice effective field theory with high-fidelity chiral interactions at N3LO.
By employing the wavefunction matching method to mitigate the sign problem and the pinhole algorithm to sample many-body correlations, we successfully reproduce the ground-state parity inversion and the extended matter radius characteristic of the halo structure.
We analyze the intrinsic density distributions and geometric shapes of $^{11}$Be in comparison with the core nucleus $^{10}$Be.
Our results reveal a prominent two-cluster structure in both nuclei and the occupation of the $\sigma$ molecular orbital by the valence neutron in $^{11}$Be.
It enhances the prolate deformation as well as the diffuse neutron tail, distinct from the $\pi$-orbital occupation observed in the $^{10}$Be ground state.}


\begin{document}

\section{Introduction}

Radioactive ion beams have opened new avenues to study nuclear structure and dynamics far away from the stability line.
Exotic phenomena such as unusually large matter distributions, later known as halo structures, were discovered in light neutron-rich nuclei~\cite{Tanihata:1988ub}.
The nucleus $^{11}$Be is one of the earliest identified and most extensively studied one-neutron halo nuclei~\cite{Fukuda:1991njn,Nakamura:1994zz}, serving as a benchmark system for understanding the properties of loosely bound quantum systems.

A striking feature of $^{11}$Be is the "parity inversion" of its ground state.
According to the standard shell model, the seventh neutron should occupy the $1p_{1/2}$ orbital, resulting in a $1/2^-$ ground state.
However, experiments established long ago that the ground state of $^{11}$Be has spin-parity $1/2^+$~\cite{Talmi:1960zz,alburger1964parity}, while the $1/2^-$ state lies at an excitation energy of 0.32~MeV.
This inversion indicates the breakdown of the N=8 magicity and the lowering of the $2s_{1/2}$ orbital relative to the $1p_{1/2}$ orbital.
The weak binding of the $s$-wave neutron leads to a spatially extended density distribution, forming the characteristic halo structure.

Experimentally, the structure of $^{11}$Be has been investigated using various probes.
Transfer reactions, such as $(d,p)$ and $(p,d)$, have been used to extract spectroscopic factors for the halo neutron.
Schmitt et al.~\cite{Schmitt:2012bt} reported spectroscopic factors of 0.71(5) for the $1/2^+$ ground state and 0.62(4) for the $1/2^-$ excited state.
Breakup and knockout reactions have also confirmed the dominant $s$-wave component of the ground state and provided information on the core excitation~\cite{Winfield:2000sq,Fortier:1999gei,Aumann:2000zz}.
Electromagnetic dissociation measurements have extracted the $B(E1)$ strength and the root-mean-square distance of the halo neutron~\cite{Fukuda:2004ct,LAND:2003tcn,Kwan:2014dha}.
Elastic scattering and breakup data have also been analyzed to constrain the core excitation and reaction mechanisms~\cite{Chen:2016jmo}.

Theoretically, reproducing the parity inversion and halo structure of $^{11}$Be has been a challenge.
Early studies highlighted the importance of a dynamic coupling between the core and the loosely bound neutron~\cite{Otsuka:1993zz}.
Approaches incorporating particle-vibration coupling~\cite{VinhMau:1995xnn,Barranco:2017rah} or core deformation~\cite{Nunes:1996zhf} have been successful in explaining the level order.
Microscopic cluster models, such as antisymmetrized molecular dynamics, have also reproduced the anomalous parity and suggested a molecular orbit picture for the valence neutron~\cite{Kanada-Enyo:2002owv,Kimura:2016fce}.
In the realm of \textit{ab initio} calculations, the no-core shell model initially faced difficulties in reproducing the parity inversion with realistic two-nucleon forces, suggesting a need for three-nucleon interactions or larger model spaces~\cite{Forssen:2004dk}.
More recent \textit{ab initio} studies using chiral interactions have emphasized the role of continuum effects and specific features of the three-nucleon force in correctly describing the spectrum~\cite{Calci:2016dfb}.
Unified descriptions using density functional theory linking parity inversion, halo, and clustering continue to be a focus of recent work~\cite{Geng:2024oex}.

Recently, a systematic \textit{ab initio} study of the beryllium isotopes from $^7$Be to $^{12}$Be was performed~\cite{Shen:2024qzi} using Nuclear Lattice Effective Field Theory (NLEFT)~\cite{Lee:2008fa,lahde2019nuclear,Lee:2025req}.
This comprehensive investigation employed high-fidelity chiral interactions up to next-to-next-to-next-to-leading order (N3LO), made possible by the Wavefunction Matching (WFM) technique~\cite{Elhatisari:2022zrb} which mitigates the Monte Carlo sign problem.
It successfully reproduced the ground state energies, radii, and electromagnetic properties of the beryllium chain, including the correct ground state of $^{11}$Be.

This proceeding will present a more detailed analysis of the $^{11}$Be halo structure as a supplement to the previous study~\cite{Shen:2024qzi}.
Specifically, we investigate the density distributions and underlying geometric structure of the ground and excited states using the pinhole algorithm.

\section{Nuclear Lattice Effective Field Theory}

In this work, we employ nuclear lattice effective field theory~\cite{Lee:2008fa,lahde2019nuclear} to investigate the structure of $^{11}$Be.
The high-fidelity \(\chi\)EFT Hamiltonian at N3LO employed in this work follows Ref.~\cite{Elhatisari:2022zrb},
\begin{equation}\label{eq:H}
    H =K + V_{\rm OPE}^{\Lambda_\pi = 300} + V_{\rm Cou.}
    + V_{\rm 2N}^{Q^4} + W_{\rm 2N}^{Q^4}
    + V_{\rm 2N,WFM}^{Q^4} + W_{\rm 2N,WFM}^{Q^4}
    + V_{\rm 3N}^{Q^3},
\end{equation}
where $K$ is the kinetic energy; $V_{\rm OPE}^{\Lambda_\pi = 300}$ is the one-pion exchange (OPE) interaction with cutoff $\Lambda_\pi = 300$~MeV;
$V_{\rm Cou.}$ is the Coulomb interaction; $V_{\rm 2N}^{Q^4}$ are the two-nucleon (2N) contact interactions at $Q^4$ chiral expansion order
and $W_{\rm 2N}^{Q^4}$ are the corresponding Galilean-invariance-restoring (GIR) terms;
$V_{\rm 2N,WFM}^{Q^4}$ and $W_{\rm 2N,WFM}^{Q^4}$ are the 2N wave function matching (WFM) interactions and GIR terms;
and $V_{\rm 3N}^{Q^3}$ are the three-nucleon (3N) contact interactions at $Q^3$ chiral expansion order.
The details on the interaction have been presented in Refs.~\cite{Li:2018ymw,Li:2019ldq,Elhatisari:2022zrb}.

To address the Monte Carlo sign problem inherent in realistic nuclear forces, we utilize the Wavefunction Matching (WFM) method~\cite{Elhatisari:2022zrb}.
This method introduces a simple Hamiltonian $H^S$ with an SU(4)-symmetric contact interaction and a smeared OPE potential.
A unitary transformation is performed on the full Hamiltonian $H$ to $H'$, so that the short-distance wave functions of $H'$ matches those of $H^S$, while the difference $H' - H^S$ is treated in first-order perturbation theory.

The ground state wave function of the simple Hamiltonian is obtained using projection Monte Carlo simulations with auxiliary fields.
The trial wave function $|\Psi_0^S\rangle$ is chosen as single Slater determinant of harmonic oscillator wave functions.
To analyze the spatial correlations and halo structure, we use the pinhole algorithm~\cite{Elhatisari:2017eno}, which samples nucleon coordinates according to the following amplitude:
\begin{equation}
    Z = \langle \Psi_0^S | M^{L_t/2} \rho(\mathbf{n}_1,\dots,\mathbf{n}_A) M^{L_t/2} | \Psi_0^S\rangle.
\end{equation}
where $M = :\exp(-a_t H_S):$ is the transfer matrix operator of the simple Hamiltonian, $a_t$ is the Euclidean time projection step, $L_t$ is the total number of time slices.
The many-body density operator $\rho(\mathbf{n}_1,\dots,\mathbf{n}_A)$ is defined as:
\begin{equation}
    \rho(\mathbf{n}_1,\dots,\mathbf{n}_A) = : a^\dagger(\mathbf{n}_1)a(\mathbf{n}_1)\dots a^\dagger(\mathbf{n}_A) a(\mathbf{n}_A):.
\end{equation}
Observables are then calculated by combining the unperturbed contributions from the sampled configurations with the perturbative corrections from the N3LO Hamiltonian~\cite{Elhatisari:2022zrb}.
See the Supplemental Materials of Ref.~\cite{Elhatisari:2017eno,Lu:2018bat} for more details of the pinhole algorithm and its pertrubative corrections.
This framework allows for the precise determination of radii and intrinsic density distributions essential for studying halo nuclei.

\section{Results and discussion}

We adopt a lattice size of $L = 10$ with lattice spacing $a = 1.32$~fm and Euclidean time lattice spacing of $a_t = 0.001$~MeV $^{-1}$.
The low-lying spectrum from $^{7}$Be to $^{12}$Be have been calculated in Ref.~\cite{Shen:2024qzi}, the energies and radii of $^{10}$Be and $^{11}$Be are summarized in Table~\ref{tab1}.
The experimental values are also listed for comparison~\cite{Wang:2021xhn,Kelley:2012qua,Nortershauser:2008vp,Tanihata:2013jwa}.
The binding energy of $^{10}$Be calculated by NLEFT is 7.2 MeV with respect to the $\alpha$-$\alpha$-neutron-neutron threshold, and 6.1 MeV with respect to the $^{9}$Be-neutron threshold.
The binding energy of the last neutron in $^{11}$Be calculated by NLEFT is 1.9 MeV with respect to the $^{10}$Be-neutron threshold, more bound comparing to the 500 keV of binding in nature.
To achieve a more accurate description of the binding energies, more efforts are needed to improve the accuracy of the ab initio calculations, such as the fine tuning of three-body interactions~\cite{Elhatisari:2025fyu}.

\begin{table}[H]
\caption{Ground state energies and radii of $^{10}$Be and $^{11}$Be obtained by the NLEFT calculation with the N3LO interaction~\cite{Shen:2024qzi}, in comparison with the experimental data~\cite{Wang:2021xhn,Kelley:2012qua,Nortershauser:2008vp,Tanihata:2013jwa}.}
\label{tab1}
\begin{tabularx}{\textwidth}{C *{6}{C}}
\toprule
\textbf{Nuclear state} &
\multicolumn{2}{c}{\textbf{Energy (MeV)}} &
\multicolumn{2}{c}{\textbf{Charge radius (fm)}} &
\multicolumn{2}{c}{\textbf{Matter radius (fm)}} \\
\cmidrule(lr){2-3} \cmidrule(lr){4-5} \cmidrule(lr){6-7}
& \textbf{NLEFT} & \textbf{Exp.}
& \textbf{NLEFT} & \textbf{Exp.}
& \textbf{NLEFT} & \textbf{Exp.} \\
\midrule
$^{10}$Be & $-63.7(3)$ & $-65.0$ & 2.51(4) & 2.36(2) & 2.53(2) & 2.30(2) \\
$^{11}$Be & $-65.6(3)$ & $-65.5$ & 2.54(4) & 2.46(2) & 2.86(1) & 2.91(5) \\
\bottomrule
\end{tabularx}
\end{table}

The ground-state parity inversion of $^{11}$Be can be reproduced by the NLEFT calculation with the N3LO interaction, and the overall agreement with the experimental data is good.
The radii of $^{10}$Be are about 10\% larger than the experimental values, while the radii of $^{11}$Be agree better.
The sudden increase of the matter radius of $^{11}$Be is manifest the halo structure.

\begin{figure}[H]
\includegraphics[width=\textwidth]{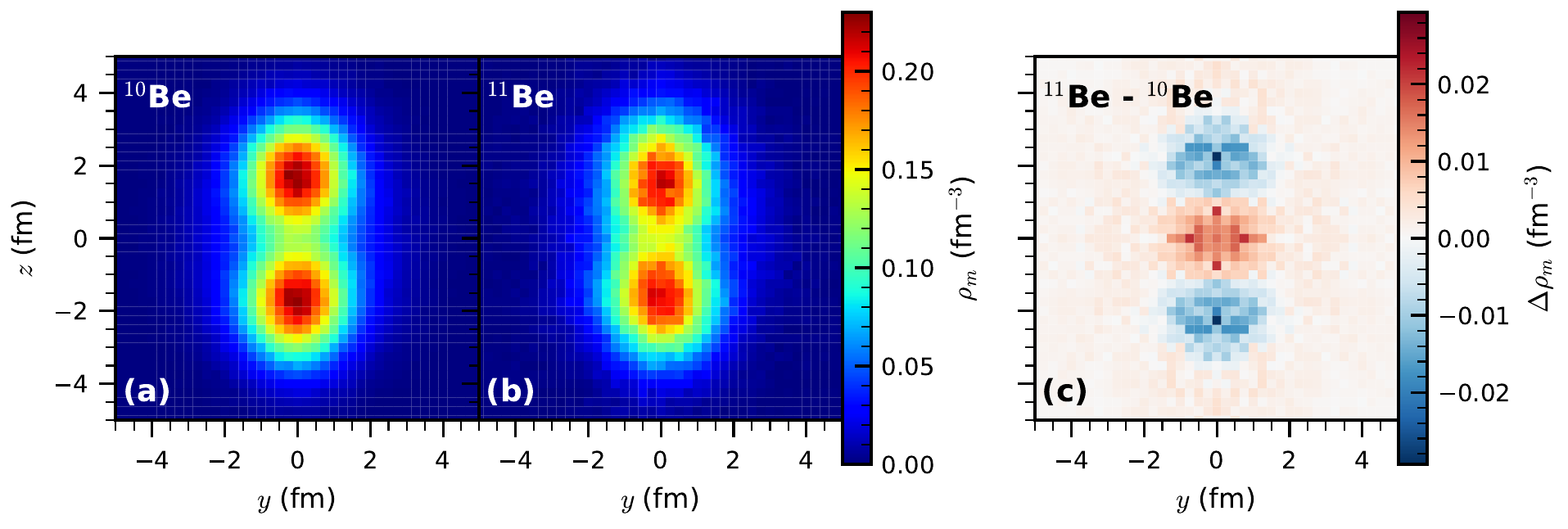}
\caption{Intrinsic 3D density distribution of (a) $^{10}$Be, (b) $^{11}$Be ground states and (c) their difference obtained by the pinhole algorithm.\label{fig1}}
\end{figure}

In Ref.~\cite{Shen:2024qzi}, the $A$-nucleons are grouped into 2 clusters composed of 2 protons and 2 neutrons that are the closest to each other according to their
spatial coordinates $\mathbf{r}_i$ sampled by the pinhole algorithm.
The extra neutrons (2 in $^{10}$Be and 3 in $^{11}$Be) are considered to be the valence neutrons.
By randomly rotating one of the clusters along the $z$-axis, the intrinsic 3D density distribution $\rho(\mathbf{r})$ is obtained.
The results of $^{10}$Be and $^{11}$Be ground states are shown in Fig.~\ref{fig1}1(a) and (b), respectively.
Their difference ($^{11}$Be - $^{10}$Be) is shown in Fig.~\ref{fig1}(c).

One can see a clear two-cluster structure in the density distribution.
Due to the different occupation of the $\pi$ and the $\sigma$ orbitals, the two nuclei show a quite different pattern of shapes.
In the $^{10}$Be ground state, the valence neutrons occupy the $\pi$ orbital surrounding  and between the two clusters~\cite{von1996two,Kanada-Enyo:1999bsw,Itagaki:1999vm,Li:2023msp,Chen:2025ung}.
The extra neutron in $^{11}$Be occupies the $\sigma$ orbital~\cite{Kanada-Enyo:1999bsw},
as can be seen in Fig.~1(c) that the center part has a higher density and to a less extent at top/bottom of the clusters.
Furthermore, the halo structure can also be seen in the difference of density distributions.

\begin{figure}[H]
\includegraphics[width=\textwidth]{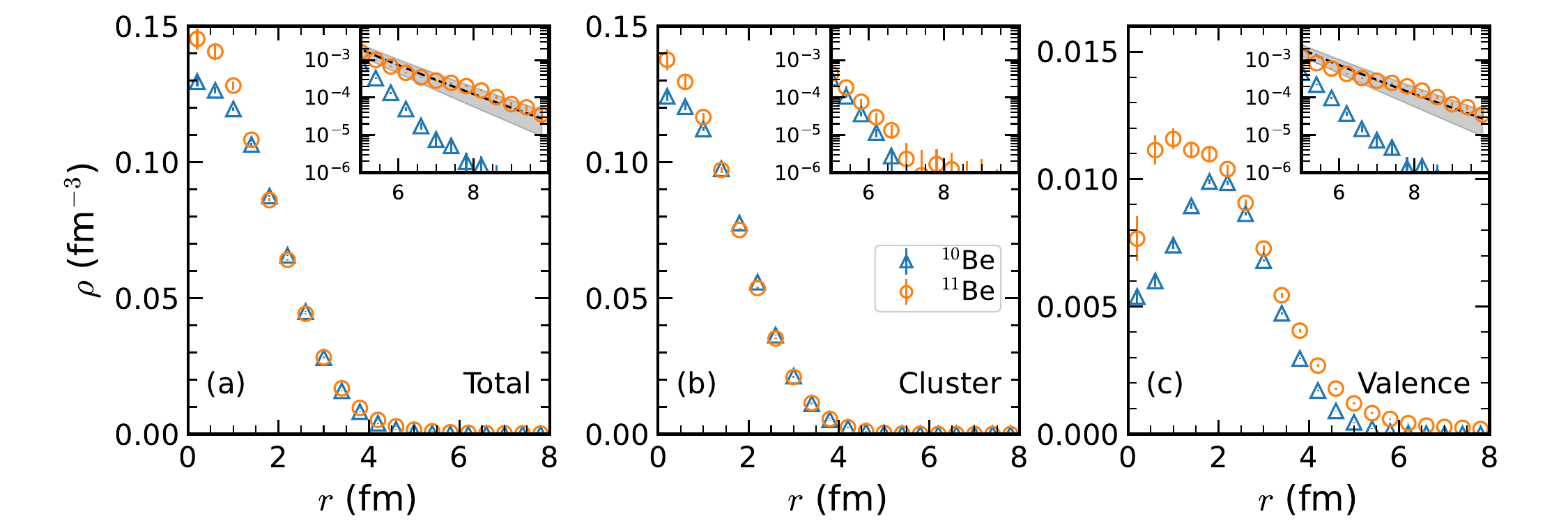}
\caption{Radial density distribution of $^{10}$Be and $^{11}$Be ground states obtained by NLEFT.
(a) Total densities, (b) densities of the 2 clusters, and (c) densities of the valence neutrons.
The dashed line in the inset indicates the asymptotic density distribution for $|E| = 1.9$ MeV, with the shadow band indicating the error propagated from the binding energy uncertainty.}\label{fig2}
\end{figure}
\unskip

To see more closely the details of the density distribution and halo character, we plot the radial density distributions of the $^{10}$Be and $^{11}$Be ground states calculated by NLEFT in Fig.~\ref{fig2}.
From left to right are total densities, densities of 2 clusters, and densities of valence neutrons.
The dashed line in Fig.~2(a) and (c) indicates the expected asymptotic density distribution proportional to $e^{-2\kappa r}/r^2$, with $\kappa = \sqrt{2mE}/\hbar \approx 0.303$ fm$^{-1}$ calculated for $|E| = 1.9$ MeV, and the shadow band indicating the error propagated from the binding energy uncertainty.
One can see at large radial distance, the density of $^{11}$Be is much more extended, showing the halo structure.
By comparing panels (a) and (c), one can see such the density at large radial distance mainly comes from the valence neutrons.
Due to the occupation of the $\sigma$ orbital, the density of $^{11}$Be near the center is larger, where both the $\sigma$-orbital valence neutron and the clusters contribute.
Due to the more prolate shape of $^{11}$Be, its cluster radial density distribution is also slightly more extended as seen from panel (b).

\begin{figure}[H]
\centering
\includegraphics[width=8.0 cm]{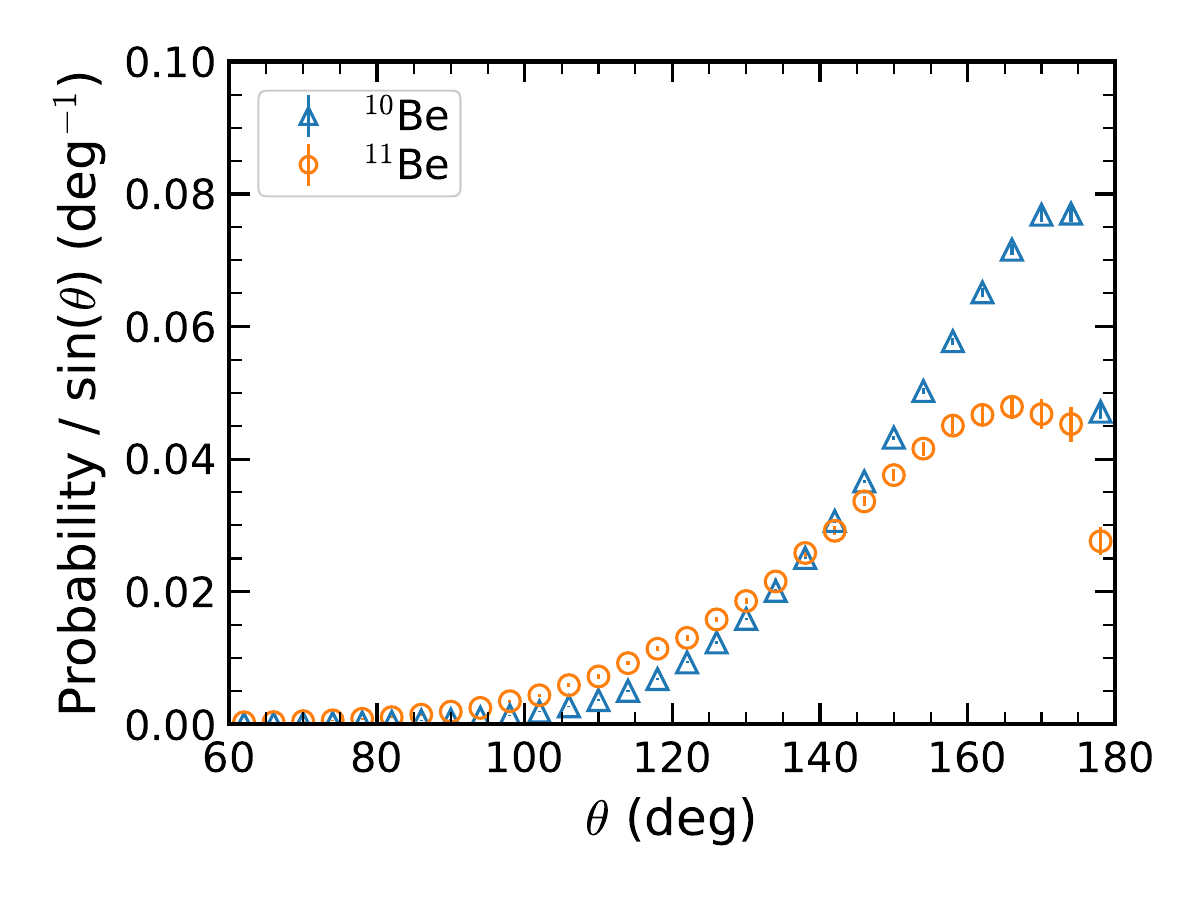}
\caption{Angular distribution of two clusters (cluster1-center-cluster2) in $^{10}$Be and $^{11}$Be ground states calculated by NLEFT.\label{fig3}}
\end{figure}   
\unskip

In Fig.~\ref{fig3}, we plot the angular distribution of two clusters (cluster1-center-cluster2) in $^{10}$Be and $^{11}$Be ground states calculated by NLEFT.
This distribution is due to the valence neutrons, as in the case of $^{8}$Be, the angle between two clusters will always be $180^\circ$.
The integration of the probability over the whole range gives one.
More valence neutrons (3 in $^{11}$Be comparing to 2 in $^{10}$Be) lead to a wider distribution of this angle.

\begin{figure}[H]
\includegraphics[width=\textwidth]{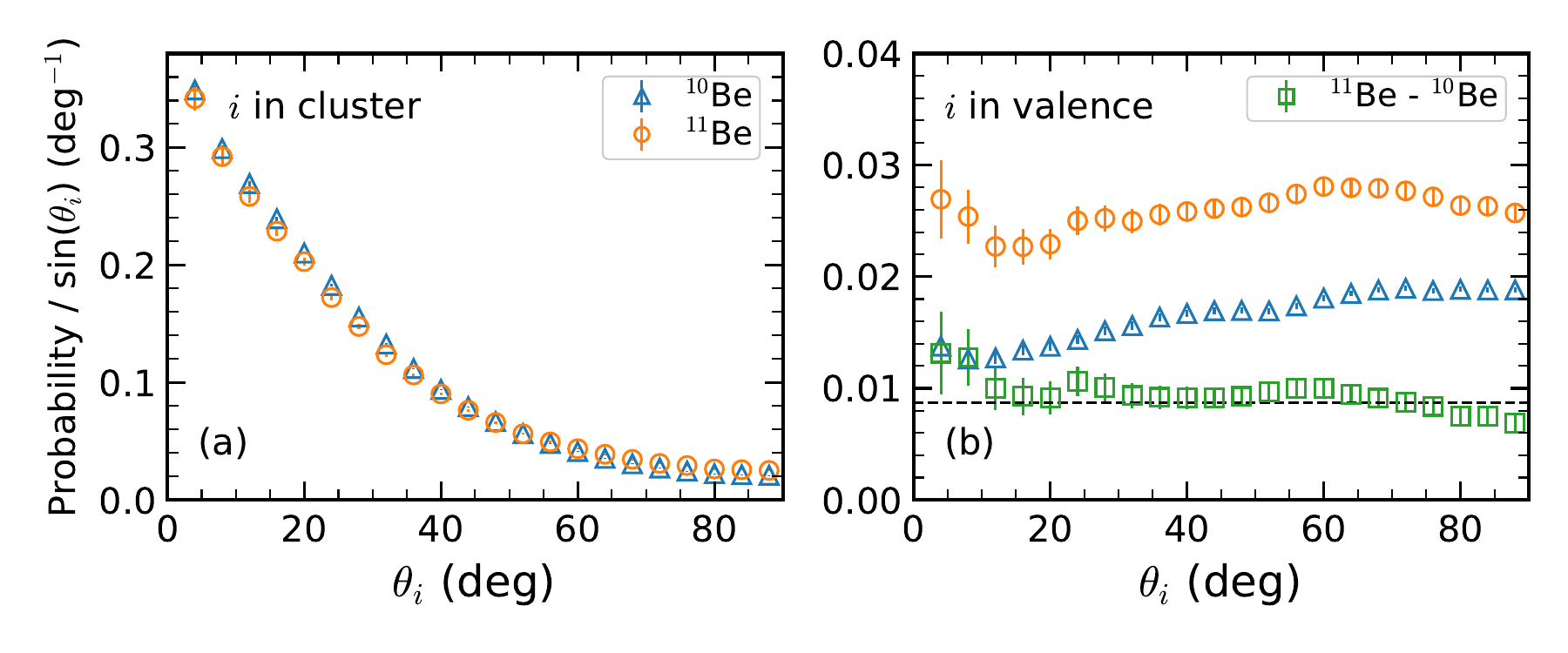}
\caption{Probability distribution of angle between nucleons and the $z$-axis in the $^{10}$Be and $^{11}$Be ground states calculated by NLEFT, in the $\alpha$-aligned frame of Fig.~\ref{fig1}.
(a) nucleons in clusters, (b)  nucleons in valence neutrons.
The gray dashed line indicates the expected angular distribution for a homogeneous sphere normalized to 1.\label{fig4}}
\end{figure}   
\unskip

In Fig.~\ref{fig4}, we plot the probability distribution of the angle between each nucleon and the $z$-axis in $^{10}$Be and $^{11}$Be ground states calculated by NLEFT, in the intrinsic frame shown in Fig.~\ref{fig1}.
The distribution is symmetric to $90^\circ$ and therefore only half is plotted.
Integration over the whole range gives the particle numbers, 8 for panel (a), 2 ($^{10}$Be) and 3 ($^{11}Be$) for panel (b).
The dashed line in Fig.~4(b) indicates the expected angular distribution for a homogeneous sphere, $p(\theta) = \frac{\pi}{360}\sin\theta$ (normalized to 1).
For nucleons in the clusters, the distribution is highest around $0^\circ$ and gradually decreases to $90^\circ$.
This is because we have aligned the clusters along the $z$-axis.
For valence neutrons, the distribution is much higher around $90^\circ$ comparing to nucleons in the clusters, as they mostly move around the neck of the two clusters.
For nucleons in the $\pi$-orbital in $^{10}$Be, while a higher probability is observed towards 90 degrees, a certain probability can also be seen at small angles.
This is related to how the one-body density is constructed from the fully-correlated many nucleon distributions.
After grouping the closest 2 protons and 2 neutrons, one cluster is rotated to the $z$-axis.
With this procedure, the valence neutron has nonvanishing probability to locate near the $z$-axis.
In panel (b), the difference between $^{10}$Be and $^{11}$Be is also plotted.
It can be seen that the extra valence neutron in $^{11}$Be leads to a higher probability around $0^\circ$, near the top of the clusters region.
Overall, this extra valence neutron distributes rather smoothly in all angles, showing both the center part of the $\sigma$ orbital and the $s$-wave character of a halo.

\section{Conclusions and Perspectives}

In this work, we have investigated the structure of the paradigmatic halo nucleus $^{11}$Be using \textit{ab initio} NLEFT calculations with high-precision N3LO chiral interactions.
By leveraging the wavefunction matching method, we overcame the sign problem and achieved a good description of the binding energies and radii for both $^{10}$Be and $^{11}$Be, properly reproducing the well-known parity inversion in the $^{11}$Be ground state.

Using the pinhole algorithm, we computed the intrinsic 3D density distributions and radial profiles, providing a microscopic view of the internal structure.
We observed well-developed cluster structures in both isotopes.
A detailed comparison of the angular and radial distributions between $^{10}$Be and $^{11}$Be demonstrates that the halo formation in $^{11}$Be is driven by the valence neutron occupying a $\sigma$-type molecular orbital.
This configuration results in a more prolate deformation and a significant extension of the neutron density at large distances compared to the $^{10}$Be core, where valence neutrons occupy $\pi$-type orbitals.

Future studies will apply this framework to explore other exotic systems near the driplines.
The successful description of clustering and halo correlations within this lattice EFT approach opens new avenues for understanding the emergence of nuclear molecular structures from first principles.

\vspace{6pt} 

\authorcontributions{
    Conceptualization, S.S. and U.G.M.; methodology, all authors; software, all authors; validation, all authors; formal analysis, S.S.; investigation, S.S.; resources, S.S. and U.G.M.; data curation, S.S. and U.G.M.; writing---original draft preparation, S.S.; writing---review and editing, all authors; visualization, S.S.; supervision, U.G.M.; project administration, U.G.M.; funding acquisition, S.S. and U.G.M.
    All authors have read and agreed to the published version of the manuscript.
}

\funding{
This work was funded by ``the Fundamental Research Funds for the Central Universities'',
National Natural Science Foundation of China under Grant No. U2541242 and 12435007.
The work of SE is supported in part by the Scientific and Technological Research Council
of Turkey (TUBITAK Project No. 123F464).
The work of UGM was supported in by the European Research Council (ERC) under the European Union’s Horizon 2020
research and innovation programme (ERC AdG EXOTIC, Grant Agreement No. 101018170),
by the CAS President’s International Fellowship Initiative (PIFI) (Grant No. 2025PD0022).
SS gratefully acknowledge the Computational resources
provided by the HPC platform of Beihang University.
SS and UGM gratefully acknowledge the Gauss Centre for
Supercomputing e.V. for funding this project by providing
computing time on the GCS Supercomputer JUWELS at
Jülich Supercomputing Centre (JSC).  DL acknowledges support from the U.S. Department of Energy grants DE-SC0026198, DE-SC0013365, and DE-SC0023175.
}

\dataavailability{The original data presented in the study are openly available in ScienceDB.}

\acknowledgments{We acknowledge the organizers of the conference “International Symposium Commemorating the 40th Anniversary of the Halo Nuclei (HALO-40)” 
for the contribution invite.}

\conflictsofinterest{The authors declare no conflict of interest.} 



\abbreviations{Abbreviations}{
The following abbreviations are used in this manuscript:
\\

\noindent 
\begin{tabular}{@{}ll}
NLEFT & Nuclear Lattice Effective Field Theory\\
$\chi$EFT & Chiral Effective Field Theory\\
WFM & Wavefunction Matching\\
GIR & Galilean Invariance Restoring\\
N3LO & Next-to-Next-to-Next-to-Leading Order\\
OPE & One-Pion Exchange\\
2N & two-nucleon\\
3N & three-nucleon\\
\end{tabular}
}

\reftitle{References}


\bibliography{be11-halo}

\PublishersNote{}
\end{document}